\documentclass[pre,aps,twocolumn,superscriptaddress,showpacs]{revtex4-1}%,showpacs
\usepackage{graphicx}
\usepackage{amsmath}

\usepackage{color}

\begin{document}

\title{New parameter-free mobility model: Opportunity priority selection model}

\author{Erjian Liu}
\affiliation{Key Laboratory of Integrated Transportation Big Data Application Technology for Transportation Industry, Beijing Jiaotong University, Beijing 100044, China}
\affiliation{Institute of Transportation System Science and Engineering,
Beijing Jiaotong University, Beijing 100044, China}

\author{Xiao-Yong Yan} \email{yanxy@bjtu.edu.cn}
\affiliation{Institute of Transportation System Science and Engineering,
Beijing Jiaotong University, Beijing 100044, China}
\affiliation{Big Data Research Center, University of Electronic Science and Technology of China, Chengdu 611731, China}

\begin{abstract}

Predicting human mobility patterns has many practical applications in urban planning, traffic engineering, infectious disease epidemiology, emergency management and location-based services. Developing a universal model capable of accurately predicting the mobility fluxes between locations is a fundamental and challenging problem in regional economics and transportation science. Here, we propose a new parameter-free model named opportunity priority selection model as an alternative in human mobility prediction. The basic assumption of the model is that an individual will select destination locations that present higher opportunity benefits than the location opportunities of the origin and the intervening opportunities between the origin and destination. We use real mobility data collected from a number of cities and countries to demonstrate the predictive ability of this simple model. The results show that the new model offers universal predictions of intracity and intercity mobility patterns that are consistent with real observations, thus suggesting that the proposed model better captures the mechanism underlying human mobility than previous models.

\end{abstract}


\maketitle

\section{Introduction}

Predicting the mobility of people, goods and information between locations is a long-term important research topic in many fields such as transportation science, economic geography, and regional economics. For more than 100 years, researchers have proposed a variety of models for predicting mobility between locations \cite{C58,S40,Z46,O15}, and these models are called spatial interaction models in economic geography \cite{SS95} and trip distribution models in transportation science \cite{OW11}.  The most widely used model is the gravity model \cite{A11} because a mobility pattern similar to Newton's law of universal gravitation is observed in many fields, such as human travel \cite{Z46,VBS06,J08,HJ16}, migration \cite{T95}, goods transportation\cite{KKGB10}, international trade \cite{F10}, mobile communications \cite{KCRB09}, and even scientific collaborations \cite{PKF12}. In this shared pattern, the mobility (or interaction) between two locations is proportional to the location masses (e.g., populations) and decays with their distance. Despite the gravity model is widely used, it relies on at least one adjustable parameter that needs to be estimated using available mobility data \cite{OW11}; moreover, the parameters of the gravity model vary from region to region \cite{B11}. Developing a universal model that can accurately predict the mobility between locations without relying on an adjustable parameter is a challenging problem.

The radiation model \cite{SGMB12} was the first-developed parameter-free mobility model, and its basic assumption is that when an individual seeks job offers from all locations, he/she will select a location that is closest to his/her home location and has higher benefits than his/her home location. The radiation model does not consider the distance as a variable and does not have any adjustable parameters. Once the spatial distribution of the population is input, the model can precisely predict the commute between locations. Another typical parameter-free mobility model is the population-weighted opportunities (PWO) model \cite{YZFDW14}, and its basic assumption is that the chance of a destination being chosen by an individual is proportional to the number of opportunities at the destination, and inversely proportional to the total population at the locations whose distances to the destination are shorter than or equal to the distance from the individual's origin to the destination. The PWO model can not only accurately predict intracity trips \cite{YZFDW14}, but can also predict intercity travel on diverse spatial scales \cite{YWGL17}. Overall, the parameter-free mobility model has become an important class of spatial interaction models \cite{BBG17}.

In this paper, we propose a new parameter-free model as an alternative in human mobility prediction. In this model, the chance of a destination being chosen by an individual is proportional to the probability that the destination has higher benefits than locations whose distances from the individual's origin are shorter than or equal to the distance from the origin to the destination. The destination selection rule of this model is very similar to the rule of the PWO model \cite{YZFDW14}, although the new model is derived from an underlying set of first principles. Although the proposed model is inspired by the model of deliberate social ties (DST) \cite{SYBS15}, our model is more suitable for practical mobility predictions, and has very broad application prospects.

\

\

\

\begin{figure*}[ht!]
\centering
{
\includegraphics[width=0.95\linewidth]{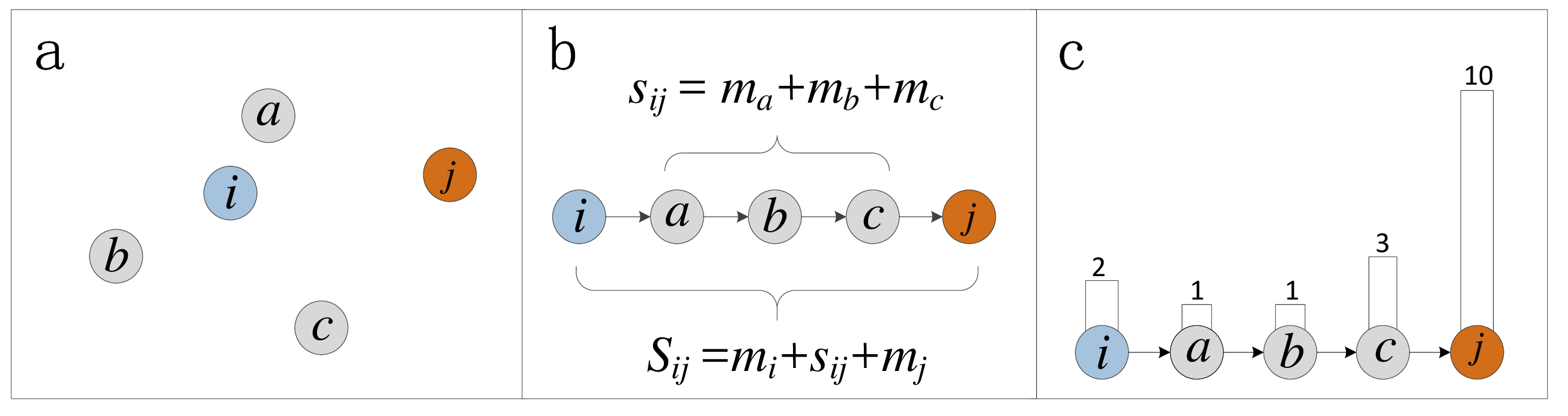}
    \caption{\label{fig-1}
    {\bf Model illustration} (a) Sketch map. Each circle represents a location. Location $i$ is the origin, and the other locations are potential destinations. (b) Intervening opportunities illustration. All locations are sorted by their distance from location $i$. If location $j$ is a destination, then the intervening opportunities are all opportunities of the locations between $i$ and $j$ such that $s_{ij} = m_a+m_b + m_c$, where $m_l$ is the number of opportunities at location $l$. (c) Model example. We use $z_l$ to represent the maximum opportunity benefit of location $l$, and bars of different heights on the figure to indicate the $z$ values for each location. If the maximum opportunity benefits of the locations are $z_i=2$, $z_a=1$, $z_b=1$, $z_c=3$, and $z_j=10$, then the individual at location $i$ will only choose location $c$ in the radiation model. However, in practice, the individual may choose $c$ or $j$ with different probabilities.}}
\end{figure*}

\section{Model}

We use the derivation of the radiation model \cite{SGMB12} as a starting point to lead our new model. In the radiation model, when an individual seeks job offers from all locations, he/she first evaluates the benefits $z$ of the employment opportunities offered by the locations. Here the number of employment opportunities in each location is proportional to the location's population, and the benefits of the opportunities are randomly chosen from a distribution $p(z)$. Then the individual will select a location that is closest to his/her home location (origin) and has a maximum  opportunity benefit that is higher than the best offer available in his/her origin.

According to the above process, for an individual at location $i$, the probability that location $j$ is closest to $i$ and has a maximum opportunity benefit that is higher than that of $i$ is  \cite{SGMB12}
\begin{equation}
\label{eq-1}
Q_{ij}=\int_0^{\infty}{\mathrm{Pr}_{m_i}(z)\mathrm{Pr}_{s_{ij}}(<z)\mathrm{Pr}_{m_{j}}(>z)} \mathrm{d}z,
\end{equation}
where $m_{i}$ is the number of opportunities at location $i$, $m_{j}$ is the number of opportunities at location $j$, {\color{black}$s_{ij}$ is the sum of the number of opportunities at all locations whose distances from $i$ are shorter than the distance from $i$ to $j$ (named {\it intervening opportunities} \cite{S40}, see Fig. \ref{fig-1}(a-b))}, $\mathrm{Pr}_{m_i}(z)$ is the probability that the maximum benefit obtained after $m_i$ samplings is exactly $z$, $\mathrm{Pr}_{s_{ij}}(<z)$ is the probability that the maximum benefit obtained after $s_{ij}$ samplings is less than $z$, and $\mathrm{Pr}_{m_{j}}(>z)$ is the probability that the maximum benefit obtained after $m_{j}$ samplings is greater than $z$.

Since $\mathrm{Pr}_{x}(<z)=p(<z)^{x}$, we can obtain 
\begin{equation}
\label{eq-2}
\mathrm{Pr}_{x}(z)=\frac{\mathrm{d}\mathrm{Pr}_{x}(<z)}{\mathrm{d}z}=x p(<z)^{x-1}\frac{\mathrm{d}p(<z)}{\mathrm{d}z}.
\end{equation}
Eq. (\ref{eq-1}) can be written as \cite{SGMB12}
\begin{equation}
\label{eq-3}
\begin{aligned}
Q_{ij}&=\int_0^{\infty}m_ip(<z)^{m_i-1}\frac{\mathrm{d}p(<z)} {\mathrm{d}z} p(<z)^{s_{ij}} [1-p(<z)^{m_{j}}]\mathrm{d}z\\
&=m_i \int_0^{1}{[p(<z)^{s_{ij}+m_{i}-1}-p(<z)^{m_{j}+s_{ij}+m_{i}-1}]}\mathrm{d}p(<z)\\
&=m_i \frac{p(<z)^{s_{ij}+m_{i}}}{s_{ij}+m_{i}}\Big\vert_0^{1}-m_i \frac{p(<z)^{m_j+s_{ij}+m_{i}}}{m_j+s_{ij}+m_{i}}\Big\vert_0^{1}\\
&=m_i \frac{1}{s_{ij}+m_{i}}-m_i \frac{1}{m_j+s_{ij}+m_{i}}\\
&=\frac{m_i m_j}{(m_i+s_{ij})(m_i+s_{ij}+m_j)}.
\end{aligned}
\end{equation}
Then, the probability that the individual at location $i$ chooses location $j$ as the destination is
\begin{equation}
\label{eq-04}
\begin{aligned}
&P_{ij}=\frac{Q_{ij}}{\sum\limits_j Q_{ij}} \\
&=\frac{m_i m_j}{(m_i+s_{ij})(m_i+s_{ij}+m_j)} / \sum\limits_{j=1}^k \frac{m_i m_j}{(m_i+s_{ij})(m_i+s_{ij}+m_j)}\\
&= \frac{m_i m_j}{(m_i+s_{ij})(m_i+s_{ij}+m_j)} / \sum\limits_{j=1}^k \left( \frac{m_i }{m_i+s_{ij}} - \frac{m_i }{m_i+s_{ij}+m_j}\right)\\
&= \frac{m_i m_j}{(m_i+s_{ij})(m_i+s_{ij}+m_j)} / \left( \frac{m_i }{m_i} - \frac{m_i }{m_i+s_{ik}+m_k}\right)\\
%&= \frac{m_i m_j}{(m_i+s_{ij})(m_i+s_{ij}+m_j)} / \left( 1- \frac{m_i }{M}\right)\\
&= \frac{1}{1 - \frac{m_i }{M}} \cdot \frac{m_i m_j}{(m_i+s_{ij})(m_i+s_{ij}+m_j)},
\end{aligned}
\end{equation}
where $k$ is the number of destinations, $M$ is the number of opportunities in all locations. This is a generalized radiation model with normalization factor for finite systems \cite{MSJB13}. Kang et al. further developed a more generalized radiation model \cite{KANG15} by imposing the normalization factor, spatial scaling exponent, searching direction and trip constraint. Since the number of opportunities in a location is assumed to be proportional to the population of the location \cite{S40,SGMB12}, the variables $m_i$,  $s_{ij}$, $m_j$ and $M$ in Eq. (\ref{eq-04}) can directly represent the populations of their corresponding locations.

The basic assumption of the radiation model is that the destination selected by the individual is the closest location whose opportunity benefits are higher than the  opportunity benefits of the origin. In practice, however, {\color{black} the individual may choose not only the closest location with higher opportunity benefits than that of origin, but also other locations with higher opportunity benefits than the benefits of origin opportunity and intervening opportunities}. Fig.  \ref{fig-1}(c) shows an example. In the radiation model, the individual at location $i$  only chooses location $c$ as the destination, although in practice, location $j$ may be selected by the individual with a higher probability. Therefore, in the new model, we assume that for the individual at location $i$, all locations whose maximum opportunity benefits are higher than the opportunity benefits of $i$ and the benefits of the intervening opportunities $s_{ij}$ can be selected as a destination. If the opportunity benefits of the locations are random variables with a distribution $p(z)$, the  probability that the maximum opportunity benefit of location $j$ will be higher than the benefits of the opportunities $m_i$ and $s_{ij}$ is
\begin{equation}
\label{eq-4}
\begin{aligned}
Q_{ij}  & = \int_0^{\infty}{\mathrm{Pr}_{m_i+s_{ij}}(z) \mathrm{Pr}_{m_{j}}(>z)}\mathrm{d}z\\
&=\int_0^{\infty} (m_i+s_{ij})p(<z)^{m_i+s_{ij}-1}\frac{\mathrm{d}p(<z)}{\mathrm{d}z} [1-p(<z)^{m_{j}}] \mathrm{d}z\\
&=(m_i+s_{ij})\int_0^{1}[p(<z)^{m_i+s_{ij}-1}-p(<z)^{m_i+s_{ij}+m_j-1}]\mathrm{d}p(<z)\\
&=(m_i+s_{ij}) \frac{p(<z)^{m_i+s_{ij}}}{m_i+s_{ij}}\Big\vert_0^{1} - (m_i+s_{ij}) \frac{p(<z)^{m_i+s_{ij}+m_{j}}}{m_i+s_{ij}+m_{j}}\Big\vert_0^{1}\\
&=1- \frac{m_i+s_{ij}}{m_i+s_{ij}+m_{j}}\\
&=\frac{m_j}{m_i+s_{ij}+m_{j}}\\
&=\frac{m_j}{S_{ij}},
\end{aligned}
\end{equation}
where $\mathrm{Pr}_{m_i+s_{ij}}(z)$ is the probability that the maximum benefit obtained after $m_i+s_{ij}$ samplings is exactly $z$, $\mathrm{Pr}_{m_{j}}(>z)$ is the probability that the maximum benefit obtained after $m_{j}$ samplings is greater than $z$, $S_{ij}$ is the sum of the intervening opportunities and the opportunities in location $i$ and $j$ such that $S_{ij} = m_i + s_{ij}+ m_j$ (see Fig.  \ref{fig-1}(b)) and the other variables have the same meanings as in Eq. (\ref{eq-3}). 

Then, the probability that the individual at location $i$ chooses location $j$ as the destination   in the new model is
\begin{equation}
\label{eq-5}
P_{ij}=\frac{Q_{ij}}{\sum\limits_j Q_{ij}} = \frac{ m_{j} / S_{ij}}{\sum\limits_j m_{j} / S_{ij}} \propto {m_{j}}/{S_{ij}}.
\end{equation}
 {\color{black} From Eqs. (\ref{eq-4}-\ref{eq-5}) we can see that the derivation procedure of the new model are consistent with that of the generalized radiation model (see Eqs. (\ref{eq-3}-\ref{eq-04})), and the two models both include the finite-size effect, i.e. the number of opportunities in all locations $M=\sum_i m_i$ is finite \cite{MSJB13}.} 
 
The probability $P_{ij}$ in Eq. (\ref{eq-5}) is very similar to the destination selection probability in the original PWO model \cite{YZFDW14}, i.e.
\begin{equation}
\label{eq-5-1}
P_{ij} = \frac{ m_{j} (1 / S_{ji}-1 /M)}{\sum\limits_j m_{j} (1 / S_{ji}-1 /M)},
\end{equation}
and that in the simplified PWO model \cite{YWGL17}, i.e.
\begin{equation}
\label{eq-5-2}
P_{ij}= \frac{ m_{j} / S_{ji}}{\sum\limits_j m_{j} / S_{ji}} \propto {m_{j}}/{S_{ji}}.
\end{equation}
{\color{black} From Eqs. (\ref{eq-5}-\ref{eq-5-2}) we can see that  the main difference between the new model and the PWO model is that the PWO model uses $S_{ji}$ as an independent variable but the new model uses $S_{ij}$.  More essential difference is that the PWO model directly establishes its destination selection rule \cite{YZFDW14,YWGL17}, whereas our new model is derived from an underlying set of initial principles}, i.e., the destination selected by an individual will be a location whose maximum opportunity benefits are not only higher than the opportunity benefits of his/her origin but also higher than the benefits of the intervening opportunities. We therefore name our new model the {\it opportunity priority selection} (OPS) model. 

\

\

\

\

\

\

\

\

\

\begin{figure*}[ht!]
\centering
{
\includegraphics[width=0.9\linewidth]{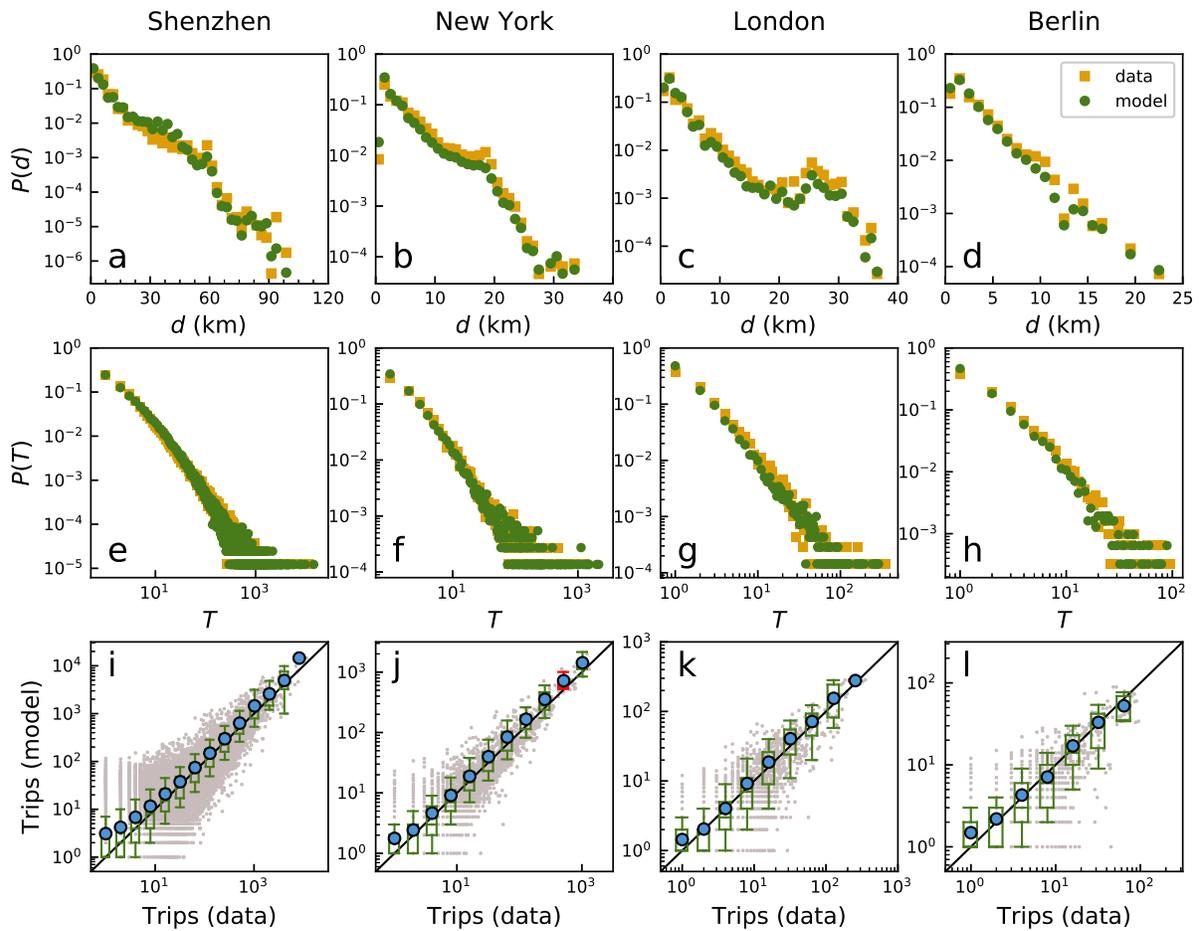}
    \caption{\label{fig-2}
    {\bf Comparison between the predictions of the  OPS model and the empirical data of intracity trips}.
(a-d) Predicted and real distributions of trip distances $P(d)$. 
(e-h) Predicted and real distributions of the number of trips $P(T)$.
(i-l) Predicted and observed trips. The gray points are the scatter plots for each pair of locations. The blue points represent the average number of predicted trips in different bins. The standard boxplots represent the distribution of the number of predicted trips in different bins of the number of observed trips.  A box is marked in green if the line $y=x$ lies between 10\% and 91\% in that bin. Otherwise, the box is marked in red.}}
\end{figure*}

If we set a separate location for each individual (i.e., there is only one individual in a location), Eq. (\ref{eq-5}) can be rewritten as $P_{ij} \propto {1}/({s_{ij}+2)}$, which is the same as the DST model without the traveling-time constraint \cite{SYBS15}. The DST model focuses on the social ties between individuals. If the attribute value of individual $j$ is higher than the attribute values of individual $i$ and the intervening opportunities $s_{ij}$, then a directed social tie from $i$ to $j$ will be built with the probability ${1}/({s_{ij}+2)}$. However, the DST model needs spatial coordinates for each individual, which are difficult to obtain in practice. In actual spatial interaction or trip distribution prediction work, researchers pay more attention to the mobility between zones (that are abstracted into locations with fixed coordinates), such as traffic analysis zones (TAZs) \cite{OW11} in urban transportation planning or cities in intercity interactions. Since data such as the populations of TAZs or cities are readily available, the zone-based mobility model is more practical.

\section{Results}

We use the intracity trip and intercity travel datasets recording the trips between different TAZs or cities to test the predictive ability of the  OPS model. This test work is equivalent to the trip distribution prediction, which is the second step of the four-step travel demand modeling process \cite{OW11}. For the  OPS model, the trip distribution prediction formula is
\begin{equation}
\label{eq-6}
T_{ij} = O_i P_{ij} =O_i \frac{ m_{j} / S_{ij}}{\sum\limits_j m_{j} / S_{ij}} ,
\end{equation}
where $T_{ij}$ is the total number of trips from origin $i$ to destination $j$, and $O_i$ is the total number of trips departed from $i$. The number of opportunities $m_j$ of location $j$ can be replaced by the total number of trips to destination $j$.

\begin{figure*}[ht!]
\centering
{
\includegraphics[width=0.9\linewidth]{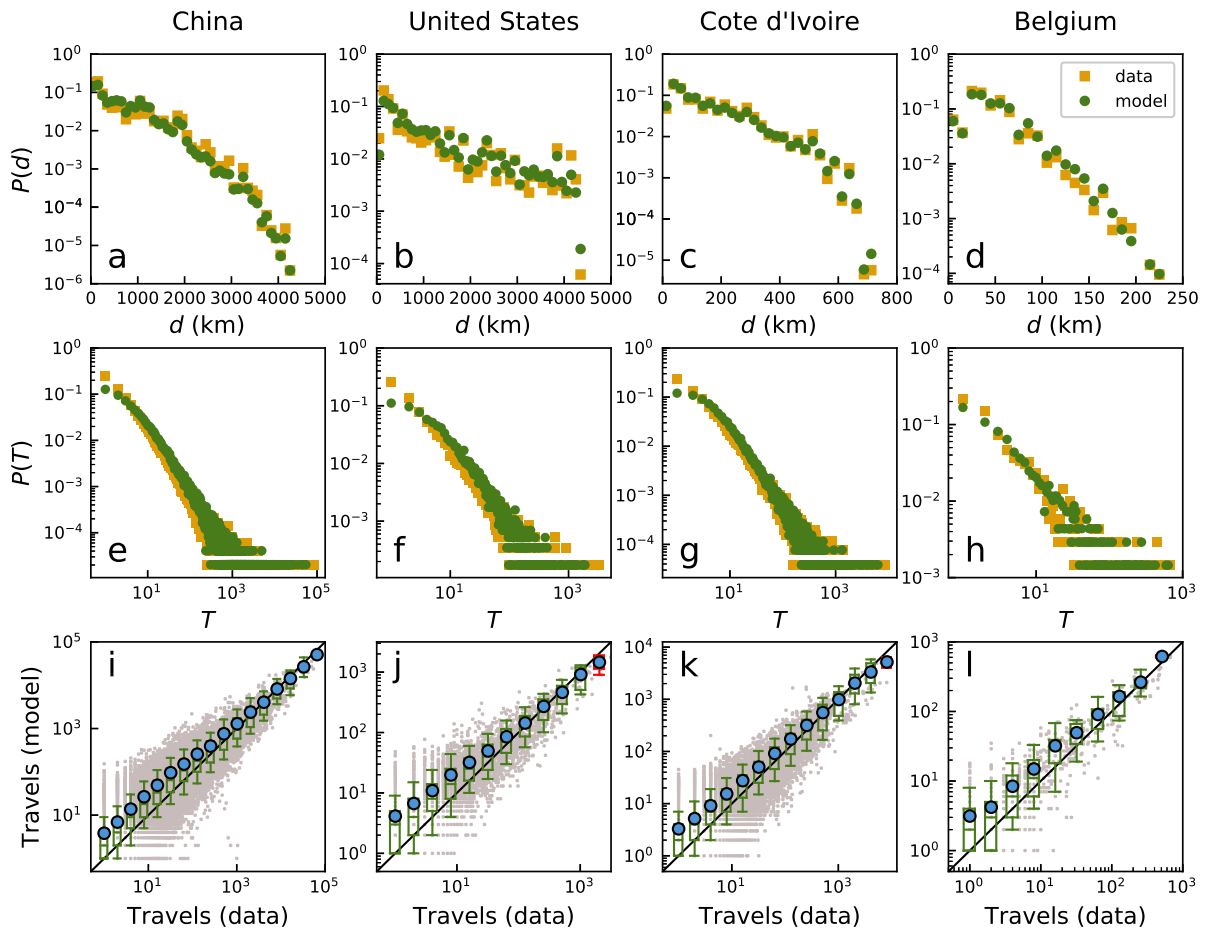}
    \caption{\label{fig-3}
    {\bf Comparison between the predictions of the  OPS model and the empirical data of intercity travel}.
(a-d) Predicted and real distributions of travel distances $P(d)$. 
(e-h) Predicted and real distributions of the number of travels $P(T)$.
(i-l) Predicted and observed travels.}}
\end{figure*}

We employ four intracity trip datasets to validate the model's predictions.
The first dataset is the trip records of taxi passengers in Shenzhen from 18 April 2011 to 26 April 2011 \cite{YZFDW14}. When a passenger gets on or gets off a taxi, the coordinates and time are recorded automatically by a GPS-based device installed in the taxi. We partition the map of Shenzhen into equal-area square TAZs, each of which is of dimension 1 km$^2$.  We extract 1070198 taxi passengers trip records from this dataset. Some evidence indicates that the average trip distance of taxi passengers is similar to the commuting distance \cite{QZW14} and the spatial distribution of taxi passengers is similar to that of populations \cite{YY14}. Thus, the taxi passengers' data can capture the intracity trip patterns to some extent \cite{YZFDW14}.
The second dataset is the check-in records of the website Foursquare~\cite{BZM12} for users in New York. Foursquare is a location-based social networking website on which users share their locations when checking in.  The dataset contains 42035 individuals, in which 23520 users have trips among different TAZs (here the TAZs are defined as the 2010 census blocks,  see www.census.gov/geo/maps-data/maps/block/2010/), and the total number of trip steps is 113279. Another two datasets are the check-in records of the website Gowalla \cite{CML11} for users in London and Berlin, which are very similar to the Foursquare dataset in New York. These geo-tagged and time-stamped check-in datasets have unique social and spatial properties that are useful for human mobility behavior studies \cite{YTAL14}.

\begin{figure*}[ht!]
\centering
{
\includegraphics[width=0.73\linewidth]{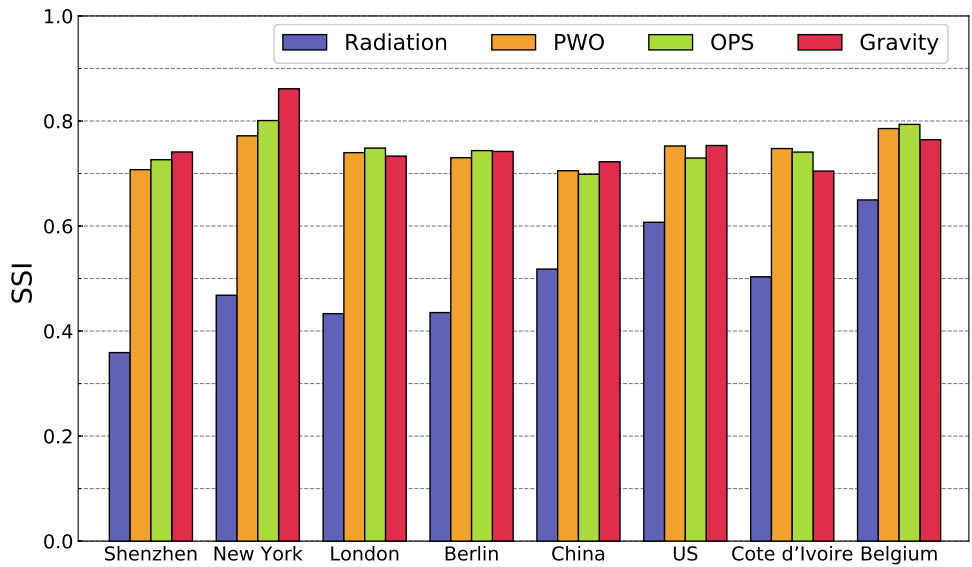}
    \caption{\label{fig-4}
    {\bf Comparison between the predictive ability of the  OPS model and the benchmark models in terms of the SSI}. The estimated parameters of the gravity model are $\beta = $ 1.20 for Shenzhen, 0.54 for New York, 1.33 for London, 1.72 for Berlin, 0.96 for China, 1.35 for US, 1.07 for Cote d'Ivoire, and 1.20 for Belgium.}}
\end{figure*}

We first investigate the trip distance distribution $P(d)$, which is a representative feature to capture human mobility behaviors \cite{GHB08,LZL12,Y13}. As shown in Fig. \ref{fig-2} (a-d), the distributions of trip distances predicted by the  OPS model are consistent with the real distributions. We next explore the distribution of the number of trips between two locations.  As shown in Fig. \ref{fig-2} (e-h), the predicted distributions $P(T)$ are consistent with the statistical results from the empirical data. A more detailed measure of a model's ability to predict mobility patterns can be implemented in terms of the trip fluxes between all pairs of locations produced by the  OPS model compared with those of real observations. As shown in Fig. \ref{fig-2} (i-l), the model's predicted and the real trip fluxes are nearly statistically indistinguishable. Overall, the  OPS model is universally applicable to intracity trip prediction.

We further use four intercity travel datasets to test the predictive ability of the  OPS model. The datasets include the check-in records of the website Sina Weibo \cite{YWGL17} for users in mainland China, the check-in records of the website Foursquare \cite{foursquaredata} for users in the continental United States, the communication records of mobile phone users in Cote d'Ivoire \cite{BECC12}, and the check-in records of the website Gowalla \cite{CML11} for users in Belgium. Among them, the Cote d'Ivoire mobile phone call detail record (CDR) dataset collects the time and positions of users making phone calls or sending text messages in a 5-month period, and the intercity social network check-in datasets have the same format with the intracity check-in datasets. Since we focus on movements among cities, all the positions within a city are regarded as the same with an identical city label. Fig. \ref{fig-3} shows that the  OPS model can also produce very good prediction results for intercity travel patterns.

We use the S{\o}rensen similarity index (SSI) \cite{S48} to compare the predictive accuracy of the mobility fluxes of the  OPS model with typical parameter-free mobility models, including the radiation model and PWO model. The SSI is a statistic tool for identifying the similarity between two samples. Here, we use a modified version \cite{L12} of the index to measure whether real fluxes are correctly reproduced (on average) by the model, and it is defined as
\begin{equation}
\label{eq-7}
\mathrm{SSI} = \frac{1}{N(N-1)}\sum^{N}_{i}{\sum^{N}_{j \neq i}{\frac{2 \min (T^{'}_{ij},T_{ij})}{T^{'}_{ij}+T_{ij}} }},
\end{equation}
where $T^{'}_{ij}$ is the number of trips from location $i$ to $j$ predicted by the model and $T_{ij}$ is the observed number of trips. Obviously, if each $T^{'}_{ij}$ is equal to $T_{ij}$, then the index is 1, whereas if all $T^{'}_{ij}$ are far from the real values, then the index is close to 0.

The comparison results are shown in Fig. \ref{fig-4}. For all studied cases, the overall prediction accuracy of the  OPS model is remarkably higher than that of the radiation model and close to that of the PWO model.  For the intracity trip distribution cases, the  OPS model is even more accurate than the PWO model.

 \begin{figure*}[ht!]
\centering
{\includegraphics[width=0.75\linewidth]{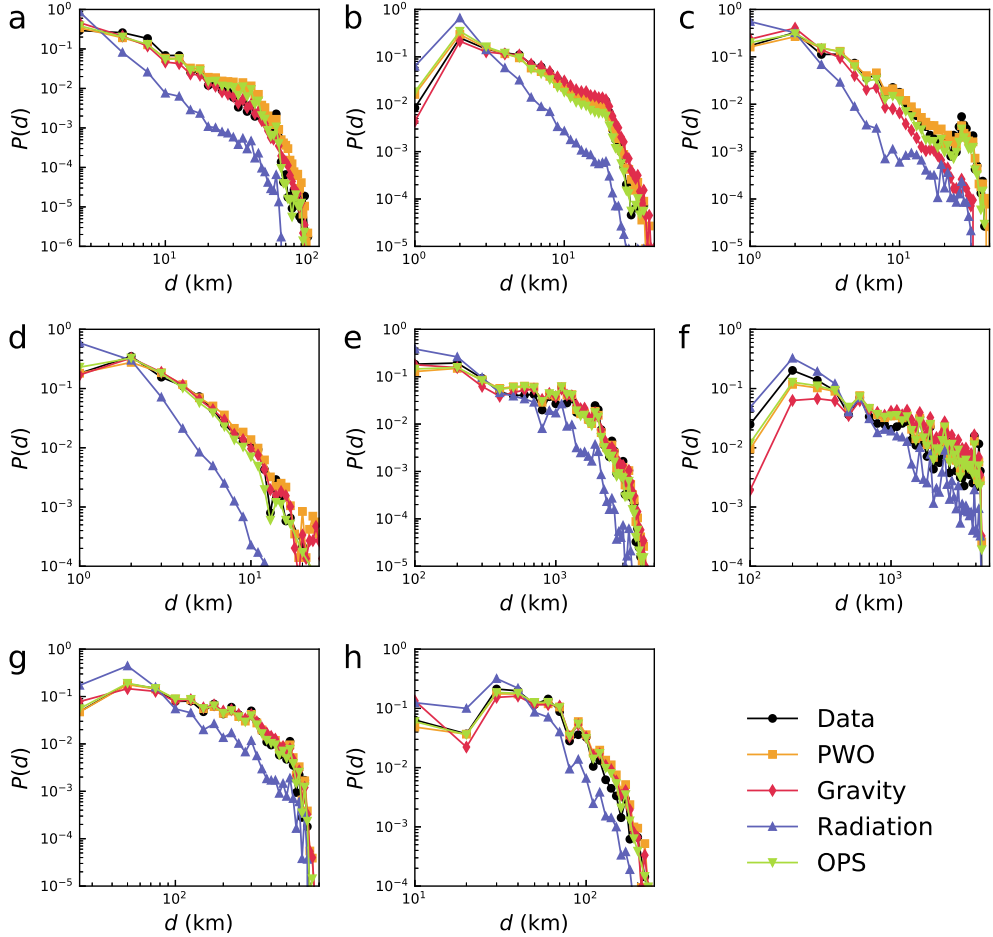}
    \caption{
    \label{fig-5}
 {\bf  Comparing the travel distance distributions predicted by the OPS model and the benchmark models. }
        (a) Shenzhen.
 (b) New York.
 (c) London.
 (d) Berlin.
(e) China.
(f) United States.
(g) Cote d'Ivoire.
(h) Belgium.
    }
    } 
\end{figure*}

We further compare the predictive accuracy of the  OPS model with the classic origin-constrained gravity model \cite{OW11}, i.e.,
\begin{equation}
\label{eq-9}
T_{ij} = O_i \frac{ m_j d_{ij}^{-\beta}}{\sum\limits_j m_j d_{ij}^{-\beta}},
\end{equation}
where $d_{ij}$ is the distance between locations $i$ and $j$, and $\beta$ is a parameter. Fig. \ref{fig-4} shows that the prediction accuracy of the  OPS model is very close to that of the gravity model and even higher than that of the gravity model in certain cases. Moreover, the gravity model has to estimate the model's parameters using real data before making a prediction, whereas the  OPS model does not require estimated parameters and can predict the trip distributions with very high accuracy, which suggests that the  OPS model is more universal than the gravity model.

Finally, we compare the travel distance distributions predicted by the OPS model and the benchmark models, as shown in Fig. \ref{fig-5}. From which we can see that the OPS model, as good as the PWO model, can precisely reproduce the observed distributions of travel distance. However, the prediction results of the radiation model deviates significantly from the real data. This implies that the assumption of the OPS model is more reasonable than that of the radiation model for general travel. In some cases (see Fig. \ref{fig-5} (c) and (f)) the results of the OPS model are even better than that of gravity model with adjustable parameters, which again implies the universality of the OPS model.

\section{Conclusions}

We developed an OPS model as an alternative to parameter-free mobility models for the prediction of intracity and intercity mobility patterns. The basic rule of this model is that the probability that a destination will be selected by an individual is proportional to the number of location opportunities at the destination and inversely proportional to the total number of the intervening opportunities and the location opportunities at the origin and destination. Compared to the PWO model, the OPS model is derived from an underlying set of first principles instead of directly making assumption about the attractiveness of locations. Compared to the radiation model, the OPS model can more reasonably reflect the actual human destination selection behavior patterns. The mobility patterns resulting from the OPS model are consistent with real data with respect to the trip distance distribution, the trip flux distribution and the average fluxes between all pairs of locations. Furthermore, the trip distribution prediction accuracy of the  OPS model is higher than that of the radiation model, similar to that of the PWO model, and closely consistent to that of the gravity model with estimated parameters, thus suggesting that our new model better captures the underlying mechanism that drives human mobility.

The main drawback of the proposed  OPS model is that it uses geographic distance as the criterion to rank the locations. The geographic distance in heterogeneous environments does not usually correspond to the actual length of travel \cite{REWGT14}. In practice, travelers use travel costs as a primary factor to evaluate which locations are more accessible. Therefore, the extended  OPS model should rank the locations according to the travel costs on the network when detailed transportation network data are available. We believe that such changes will definitely increase the accuracy of the  OPS model predictions.

Human mobility behavior is strongly correlated with social interactions \cite{CML11,FLHRZ17}. The  OPS model can not only predict human mobility patterns, but can also evaluate the social tie connectivity \cite{SYBS15} of cities and countries. Since social tie density is a function of population density \cite{PGKCP13}, the  OPS model can measure the overall social tie connectivity and local connectivity of each location using population distribution data. Moreover, the  OPS model can be used to quantify and assess the impacts of demographic changes, transportation networks and other infrastructure developments. Thus, it has broad application prospects in urban planning, transportation management, infrastructure assessment, etc.

\

{\bf Acknowledgements:}
X.Y. was supported by NSFC under Grant Nos. 71822102, 71621001 and 71671015.

\end{document}